\documentclass[twocolumn,10pt]{article} % here we use the article class, rather than elsarticle

\usepackage[square,numbers,sort&compress,comma]{natbib}

\usepackage{amsmath}
\usepackage{amssymb}
\usepackage{caption}
\usepackage{graphicx}
\usepackage{latexsym}
\usepackage{times}
\usepackage[pagewise]{lineno}
\usepackage{lipsum}

\usepackage[version=4]{mhchem}
\usepackage[per-mode=symbol]{siunitx}
\usepackage{subcaption}
\topmargin - 12pt % might need to be set to 0pt for some installations
\renewenvironment{abstract}%
              {% - begin definition
               \small% - select font
               {\bfseries \abstractname}% - select font
               \par% - end a paragraph (skip \parsep)
               \vspace{10pt}% - add vertical space
              }% - complete definition

\renewcommand\abstractname{Abstract}

\newcommand{\nomenclature}% - name of command
              [1]% - number of arguments
              {% - begin definition
               \bgroup% - begin a local group
               \flushleft% - turn on flushleft option
               \small\bf% - select font
               #1% - insert title text
               \par% - end a paragraph (skip \parsep)
               \egroup% - terminate local group
              }% - complete definition

\renewcommand{\section}% - name of command
              [1]% - number of arguments
              {% - begin definition
               \bgroup% - begin a local group
               \flushleft% - turn on flushleft option
               \small\bf% - select font
               \refstepcounter{section}% - increment counter
               \arabic{section}. #1% - insert title text
               \par% - end a paragraph (skip \parsep)
               \egroup% - terminate local group
              }% - complete definition

\renewcommand{\subsection}% - name of command
              [1]% - number of arguments
              {% - begin definition
               \bgroup% - begin a local group
               \flushleft% - turn on flushleft option
               \small\em% - select font
               \refstepcounter{subsection}% - increment counter
               \arabic{section}.% - insert title text
               \arabic{subsection}. #1% - insert title text
               \par% - end a paragraph (skip \parsep)
               \egroup% - terminate local group
              }% - complete definition

\renewcommand{\subsubsection}% - name of command
              [1]% - number of arguments
              {% - begin definition
               \bgroup% - begin a local group
               \flushleft% - turn on flushleft option
               \small\em% - select font
               \refstepcounter{subsubsection}% - increment counter
               \arabic{section}.% - insert title text
               \arabic{subsection}.% - insert title text
               \arabic{subsubsection}. #1% - insert title text
               \par% - end a paragraph (skip \parsep)
               \egroup% - terminate local group
              }% - complete definition

  \newcommand{\acknowledgement}% - name of command
              [1]% - number of arguments
              {% - begin definition
               \bgroup% - begin a local group
               \flushleft% - turn on flushleft option
               \small\bf% - select font
               #1% - insert title text
               \par% - end a paragraph (skip \parsep)
               \egroup% - terminate local group
              }% - complete definition

  \newcommand{\sectionbib}% - name of command
              [1]% - number of arguments
              {% - begin definition
               \bgroup% - begin a local group
               \flushleft% - turn on flushleft option
               \small\bf% - select font
               #1% - insert title text
               \par% - end a paragraph (skip \parsep)
               \egroup% - terminate local group
              }% - complete definition

\begin{document}

\title{\LARGE Analysis of soot formation in a lab-scale Rich-Quench-Lean combustor using LES with tabulated chemistry}

\author{{\large Leonardo Pachano$^{a,b,*}$, Abhijit Kalbhor$^{c}$, Daniel Mira$^{a}$, Jeroen van Oijen$^{c,d}$}\\[10pt]
        {\footnotesize \em $^a$Barcelona Supercomputing Center (BSC), Plaça Eusebi Güell, 1-3 08034, Barcelona, Spain}\\[-5pt]
        {\footnotesize \em $^b$Universitat Politècnica de València (UPV), Valencia 46022, Spain}\\[-5pt]
        {\footnotesize \em $^c$Department of Mechanical Engineering, Eindhoven University of Technology, 5600 MB, Eindhoven, the Netherlands}\\[-5pt]
        {\footnotesize \em $^d$Eindhoven Institute for Renewable Energy Systems (EIRES), 5600 MB, Eindhoven, the Netherlands}}

\date{}

% -------------------------------------------------------------------- %
% -------------------------------------------------------------------- %
% -------------------------------------------------------------------- %

\small
\baselineskip 10pt

% -------------------------------------------------------------------- %
% -------------------------------------------------------------------- %
% -------------------------------------------------------------------- %

\twocolumn[\begin{@twocolumnfalse}
\vspace{50pt}
\maketitle
\vspace{40pt}
\rule{\textwidth}{0.5pt}
\begin{abstract} % 100 to 300 words.

A numerical study of the effects of air dilution on soot formation and particle dynamics in a lab-scale rich-quench-lean (RQL) combustor is presented using large-eddy simulations (LES) with tabulated chemistry.
The modelling approach comprises a flamelet generated manifold (FGM) turbulent combustion model and an efficient discrete sectional method with clustering of sections (CDSM) to model soot formation. Three operating conditions are studied including a reference case without secondary air dilution and two cases with varying air dilution
levels.
For the latter, the split between primary and dilution air is varied from 80:20 to 60:40 expressed in percentage values. 
The study aims to investigate the effect of air dilution on the formation and oxidation of soot for various air splits using a
direct comparison with the available experimental data.
The results show a good correlation between predicted flame topology and spatial distribution of the soot volume fraction with the experimental observations. The introduction of air dilution is found to limit the production of soot with a more drastic reduction for the 40\% dilution case compared to the 20\%  condition. Predicted particle size distributions (PSD) from the case without secondary air dilution correlate well with scanning mobility particle sizer (SMPS) measurements although fewer and smaller particles are predicted with air dilution. Leaner mixtures and enhanced oxidation, resulting from the interaction with air dilution jets, favor the decrease in soot formation.

\end{abstract}
\vspace{10pt}
\parbox{1.0\textwidth}{\footnotesize {\em Keywords:} Large-eddy simulation; Soot modelling; Discrete sectional method; Tabulated chemistry; Rich-quench-lean combustor}
\rule{\textwidth}{0.5pt}
\vspace{10pt}
\end{@twocolumnfalse}] 

\clearpage

\section{Introduction} \addvspace{10pt}

Undesired soot emissions from combustion systems in the transport and energy sectors have a detrimental impact on the environment and human health. Consequently, extensive research efforts within the combustion community aim to mitigate soot emissions, especially in aviation transport due to ever-stricter regulations. 
The rich-quench-lean (RQL) combustion mode is prevalent in civil aviation ~\cite{samuelsen2006} and is analysed in this study.
RQL combustors are characterized by an initial fuel-rich zone, followed by quick mixing (facilitated through air dilution jets) leading to a subsequent lean zone.
Initially developed to limit \ce{NO_x} emissions, RQL combustors have also been studied to understand soot formation dynamics~\cite{geigle2017,elhelou2021,defalco2021}. The Cambridge RQL (C-RQL) combustor~\cite{elhelou2021,defalco2021} has been experimentally characterized to investigate soot formation under diverse operating conditions. This lab-scale gas turbine combustor model allows for a comprehensive study on the influence of air dilution on the production of soot achieved by varying the split between primary and dilution air and the location at which air dilution jets are introduced.
The experimental database from the C-RQL combustor comprises scanning mobility particle sizer (SMPS) particle size distribution (PSD) measurements providing valuable insights into PSD evolution, particularly in turbulent flames where experimental results are scarce~\cite{lindstedt2023}.

Although direct numerical simulations (DNS)~\cite{lignell2008,attili2014} enable detailed insight into soot formation, large-eddy simulations (LES) offers a computationally affordable alternative to study practical applications~\cite{lindstedt2023}.
Soot calculations using LES have been successfully conducted on RQL systems. The pressurized RQL combustor from the German Aerospace Center (DLR)~\cite{geigle2015a} has been the subject of numerous investigations using simplified approaches like two-equation soot models \cite{felden2018, eberle2018}, but also more detailed approaches based on the method of moments \cite{chong2018,cokuslu2022} or the sectional method \cite{grader2018,garcia-oliver2024}. The state-of-the-art on the C-RQL comprises the works of Giusti et al.~\cite{giusti2018} and Gkantonas et al.~\cite{gkantonas2020} through LES coupled with the conditional moment closure (CMC). In these works, various splits of primary and dilution air with two global equivalence ratios were conducted. A two-equation soot model was used in \cite{giusti2018}, while a detailed sectional approach was used in \cite{gkantonas2020}. These works exposed the intricate nature of the soot production process in the C-RQL combustor, emphasizing the tight coupling between mixing and combustion processes, and the strong interaction with the air dilution jets.

This work aligns with these efforts in order to investigate the influence of air dilution on soot formation for lean operation of the C-RQL combustor.
Three operating conditions are considered: a reference case without secondary air dilution and two cases with varying air dilution (20\% and 40\% of the total air supply), while maintaining a constant global equivalence ratio $\phi_g$ = 0.3. The modelling strategy integrates flamelet generated manifold (FGM) chemistry and a computationally efficient discrete sectional method with clustering of sections (CDSM) for soot modelling~\cite{kalbhor2023}. The objectives of the work are twofold:
(1) to evaluate the predictability of the LES FGM-CDSM approach through a direct comparison with experimental results for the mean flame topology, spatial distribution of soot volume fraction ($f_v$), and the PSD evolution,
(2) to provide insights on the influence of air dilution through side jets on soot formation and the resulting PSD.
This is, to the best of the authors' knowledge, the first numerical investigation of the C-RQL experiments published in \cite{elhelou2021,defalco2021} providing a systematic study on the air split ratio.
The numerical methodology is based upon the prior work of the authors on the clustering method (CDSM)~\cite{kalbhor2023} and the evaluation of the LES FGM-DSM approach in turbulent non-premixed jet flames~\cite{kalbhor2024}.
The application of the LES FGM-CDSM approach, recently evaluated
for the DLR RQL combustor~\cite{garcia-oliver2024}, is extended in this work to the C-RQL combustor to demonstrate its robustness and applicability across different combustion systems and conditions.

%----------------------------------------------------------------
\section{Case of study and numerical setup} \addvspace{10pt}

\begin{figure}[ht]
\centering
\includegraphics[width=0.85\linewidth]{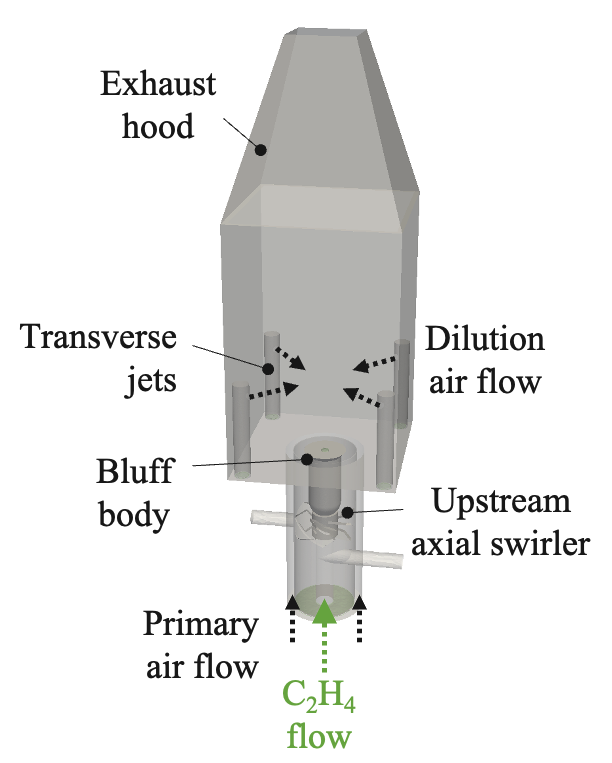}
\caption{Computational domain used in the LES.}
\label{fig:layout}
\end{figure}

The case of study is a lab-scale gas turbine combustor operating under the RQL concept. The combustion process is characterized by a bluff body, swirl-stabilized non-premixed ethylene/air flame. The combustor operates at atmospheric pressure with a central fuel inlet with \SI{4}{mm} diameter and an axial swirler that supplies ambient temperature primary air into the system. The primary air inlet is an annulus formed by a bluff body of \SI{25}{mm} diameter and an outer pipe wall of \SI{37}{mm} diameter.
The combustor has a square cross section of 97\! $\times$\! \SI{97}{mm^2} and a length of \SI{150}{mm}. Dilution air can be supplied through four tubes at the corners of the combustion chamber. Fig.~\ref{fig:layout} shows an overview of the computational domain used for the LES in this work. The air split ratio between primary and dilution air, as well as the location of the dilution jets, can be adjusted.

In this work, three operating conditions, given in Table~\ref{tab:bound_conditions}, are investigated for varied bulk inlet velocity of primary air ($U_{p,a}$) and dilution air ($U_{d,a}$). 
Following the nomenclature from~\cite{elhelou2021,defalco2021}, the Baseline case denotes a case without secondary air dilution, Case III involves an air split of 80\% primary air and 20\% dilution air, while Case IV features an air split ratio of 60\% primary air and 40\% dilution air. 
The height at which the dilution air is injected, fuel inlet velocity ($U_f$), and $\phi_g$ are kept constant at \SI{47}{mm}, \SI{15}{m/s}, and 0.3, respectively. 

\begin{figure}[ht]
\centering
\includegraphics[width=0.9\linewidth]{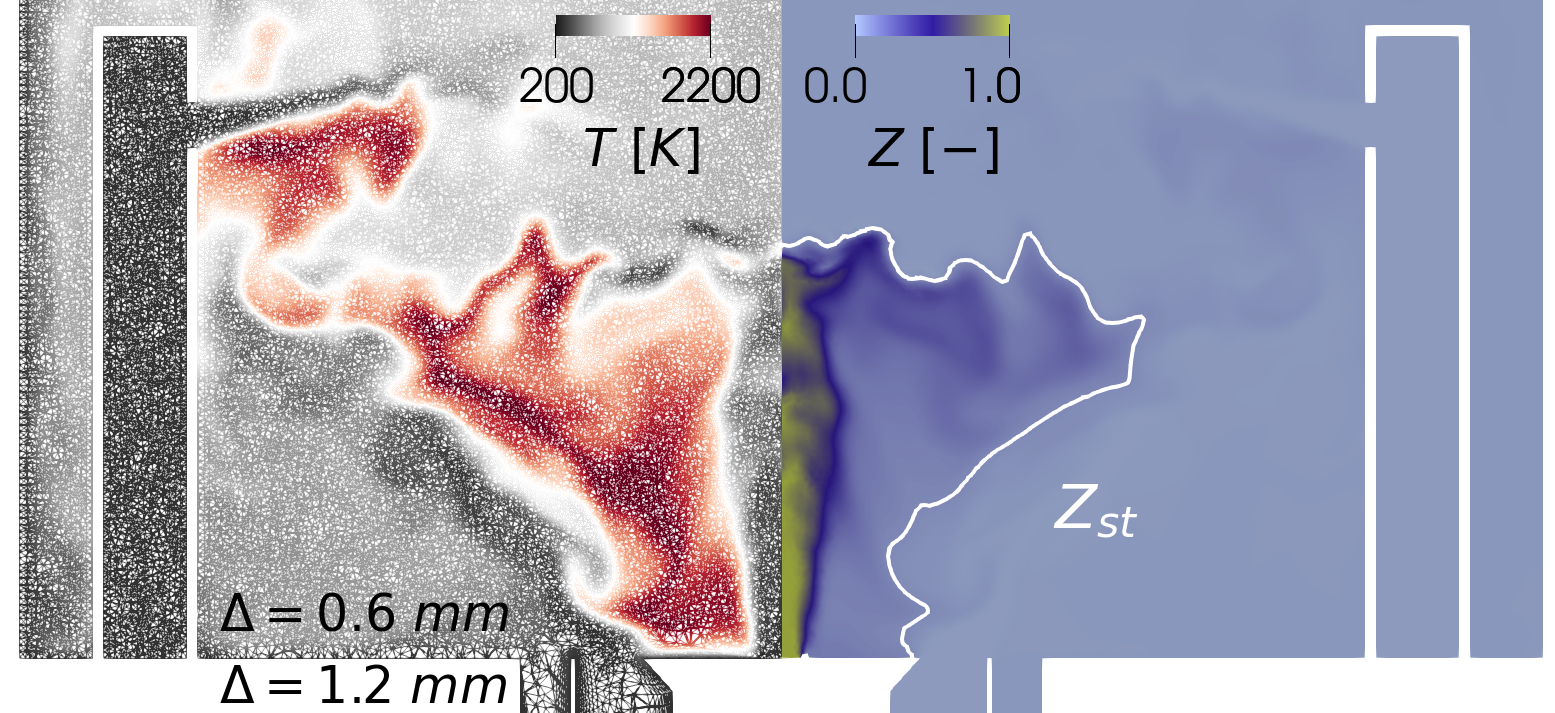}
\caption{A sector of the combustor chamber showing the mesh refinement colored by instantaneous temperature (left) and mixture fraction, $Z$ (right).}
\label{fig:mesh}
\end{figure}

\begin{table}[h] \small
\caption{Operating conditions.}
\centerline{\begin{tabular}{lccc}
\hline
\text{Case}     &  $U_{p,a}$ [m/s]     & $U_{d,a}$ [m/s]& Air split \\ \hline
Baseline             & 15.8          & -             & 100:0 \\
Case III             & 12.7          & 40            & 80:20 \\ 
Case IV              & 9.64          & 76            & 60:40 \\ \hline
\end{tabular}}
\label{tab:bound_conditions}
\end{table}

A standard convective out-flow and a steady in-flow boundary conditions are used at the outlet and the primary air inlet of the combustor, respectively. Synthetic turbulence \cite{kempf2005} is introduced in the boundary condition at the fuel inlet and the inlets of dilution air. 
The domain is discretized with a hybrid mesh comprising 10 million degrees of freedom. A sector of the combustor model is shown in Fig.~\ref{fig:mesh}. The mesh is colored by instantaneous temperature (Fig.~\ref{fig:mesh} (left)) exhibiting the interaction of cold dilution air and hot gases from the flame. The minimum element size is \SI{0.6}{mm} in the combustion chamber and \SI{1.2}{mm} elsewhere. Fig.~\ref{fig:mesh} (right) depicts incoming fuel jet from the instantaneous $Z$ field. Heat losses are accounted for through two mechanisms. First, a radiative heat loss source term, included in the energy transport equation, is used to account for gas-phase radiation with the optically thin approach. Second, isothermal wall boundary conditions are imposed introducing convective heat losses to the walls.

The code Alya~\cite{vazquez2016} from the Barcelona Supercomputing Center (BSC) has integrated the proposed LES-FGM-CDSM framework and it is employed here for the calculations. The spatial discretization is based on a low-dissipation conservative finite element scheme for low-Mach number reacting flows~\cite{both2020}, in which a non-incremental fractional-step approach, modified for variable density flows, is used for the stabilization of the continuity equation. Scalar transport is solved with the Algebraic Subgrid Scale (ASGS) method ~\cite{both2020}. A time integration third order Runge-Kutta scheme is employed for momentum and scalars.
Statistics were collected over three flow-through times. Mean results presented in the following sections are first time-averaged and then azimuthal-averaged with a resolution of 1 degree.

%---------------------------------------------------------------
\section{Modelling approach} \addvspace{10pt}

The modelling approach is based on the coupling of flamelet generated manifold (FGM) chemistry~\cite{vanoijen2016}
with a discrete sectional method (DSM)~\cite{hoerlle2019} using the clustering of sections (denoted as FGM-CDSM) introduced by Kalbhor et al.~\cite{kalbhor2023} to model soot formation. 
The governing equations for LES correspond to the filtered Navier-Stokes equations in the low-Mach number limit using the enthalpy equation with the assumption of unity Lewis number \cite{both2020}.
Unresolved heat fluxes and momentum transport are modeled through a gradient approach~\cite{miramartinez2014} and the Boussinesq approximation~\cite{poinsot2005}, respectively. Vreman's model ~\cite{vreman2004}, with  $c_k$ = 0.1, is used to estimate the eddy viscosity ($\nu_t$) as in previous works \cite{mira2020,kalbhor2024}.

\subsection{Turbulent combustion modelling} \addvspace{10pt}

The laminar flamelet database for the FGM-CDSM approach is constructed
from the tabulation of 1D counterflow diffusion flames using the
code CHEM1D \cite{somers1994}.
Gas-phase chemistry is described based on the chemical kinetics scheme KM2 from Wang et al. \cite{wang2013}. Flamelet solutions at different enthalpy levels are considered to account for the effect of heat loss. The manifold is then described by three control variables including Bilger's mixture fraction ($Z$), a reaction progress variable ($Y_c$), and a scaled enthalpy ($\mathcal{H}$). A linear combination of the mass fractions of \ce{H2O}, \ce{CO}, \ce{H2}, \ce{CO2}, \ce{C2H2}, and \ce{A}4 (pyrene) is used to define $Y_c$, while $\mathcal{H}$ is scaled with the minimum and maximum enthalpy levels in the flamelet for each value of $Z$. Filtered transport equations for the two control variables $Z$ and $Y_c$ are used to recover the thermochemical states at runtime using a gradient diffusion approach for the unresolved scalar transport~\cite{both2020,kalbhor2024}.

Subgrid scale turbulence-chemistry interactions are accounted for through a presumed probability density function (PDF) approach. A $\beta$-PDF function is used to describe the statistical distribution of $Z$ as these marginal PDFs are shown to be suitable for the convolution of flamelet solutions in other LES studies~\cite{domingo2008, massey2021}. The subgrid variance of mixture fraction is defined as $Z_v\!=\!\widetilde{Z^2}-\widetilde{Z}\widetilde{Z}$ and it is obtained from the solution of a transport equation where the unresolved contribution to the scalar dissipation rate is modeled assuming linear relaxation at sub-grid level \cite{domingo2008}. A $\delta$-function is used for both $Y_c$ and $\mathcal{H}$. Details of the modelling approach can be found in a previous work~\cite{kalbhor2024}.
The turbulent database is discretized with 101$\times$11$\times$101$\times$21 points for $Z\!\times \!Z_v\!\times\! Y_c  \!\times\! \mathcal{H}$ space, respectively.

\subsection{Soot modelling} \addvspace{10pt}

FGM-DSM has been applied to LES of a turbulent non-premixed jet flame~\cite{kalbhor2024}, while the FGM-CDSM formulation has been recently validated in laminar non-premixed flames \cite{kalbhor2023}. In this section, the formalism of the FGM-CDSM approach in the context of LES is briefly described. Following the DSM approach, the population of soot particles is discretized (based on volume) in a given number of sections $n_{sec}$ (set to thirty here). 
At flamelet level, the soot mass fraction ($Y_{s,i}$) in section $i$ is transported by accounting for soot convection, thermophoresis, diffusion, and complete soot kinetics. The soot chemistry accounts for the contribution of nucleation based on \ce{A}4 dimerization, PAH condensation based on the collisions of soot particles and \ce{A}4 molecules, coagulation based on \cite{kumar1996}, and surface chemistry (growth and oxidation of particles) through the \ce{H}-Abstraction \ce{C2H2}-Addition (HACA) mechanism \cite{frenklach1991,appel2000}. Within the HACA mechanism, oxidation accounts for the interaction of soot particles with \ce{O2} and \ce{OH} with collision efficiency ($\gamma_{\ce{OH}}$) of 0.13.
Soot particles are considered as spheres neglecting morphological properties for simplicity. The consumption of soot precursors is accounted for during the 1D flamelet calculation using two way coupling between gas and solid phase. A more detailed description of the soot model and sub-processes along with validation results can be found in \cite{hoerlle2019}.

Following the approach presented in \cite{kalbhor2023}, soot sections are reduced into $n_{clt}$ clusters (six in this work) assuming that within a given cluster the soot PSD from the original sections is preserved.
Under this assumption, the soot mass fraction $Y_{s,j}^{clt}$ in cluster $j$ is computed as $Y_{s,j}^{clt} = \sum_{i=i_{j}^\mathrm{min}}^{i_{j}^\mathrm{max}} Y_{s,i}$, where the lower and upper limit of the clustered sections $i$ are denoted as $i_{j}^\mathrm{min}$ and $i_{j}^\mathrm{max}$, respectively. By combining thirty sections into six clusters, the computational cost is substantially reduced as only six new transport equations are needed to account for soot formation in the LES. The filtered transport equation for the soot mass fraction in cluster $j$ follows Eq.~\eqref{eq:soot_alya}:
\begin{align}
    \begin{split}
    \label{eq:soot_alya}
    \frac{\partial \left(\bar{\rho} \widetilde{Y}_{s,j}^{clt} \right)}{\partial t} + &\nabla \cdot \left(\bar{\rho} \left[ \boldsymbol{\widetilde{u}} + \boldsymbol{\widetilde{v}}_T \right] \widetilde{Y}_{s,j}^{clt} \right) =\\
    & \nabla \cdot \left[\bar{\rho} \left( \bar{D}_s + \frac{\nu_t}{Sc_t} \right) \nabla \widetilde{Y}_{s,j}^{clt} \right]\\
    & + \overline{\dot{\omega}}_{s,j}^{clt} \ \ \ \ \ \forall \ j \in \left[1, n_{clt} \right],
    \end{split}
\end{align}
\noindent where the filtered thermophoretic velocity is $\boldsymbol{\widetilde{v}}_T\!=\!-0.554 \bar{\nu} (\nabla \widetilde{T} / \widetilde{T})$ and $\overline{\dot{\omega}}_{s,j}^{clt}$ is the filtered source term in cluster $j$. Following the presumed PDF approach described for gas-phase, $\overline{\dot{\omega}}_{s,j}^{clt}$ is defined as:
\begin{align}
    \begin{split}
    \label{eq:soot_pdf}
    \overline{\dot{\omega}}_{s,j}^{clt} = \bar{\rho} \iint\frac{1}{\rho} \dot{\omega}_{s,j}^{clt} 
    \widetilde{P}\left(\phi_g,\phi_s\right) d\phi_g d\phi_s.
    \end{split}
\end{align}

In Eq.~\eqref{eq:soot_pdf}, $\phi_g$ denotes gas-phase thermo-chemical variables, while $\phi_s$ denotes soot variables (e.g., number density). The joint PDF is treated as $\widetilde{P}(\phi_g,\phi_s)\!=\!\widetilde{P}(\phi_g)P(\phi_s|\phi_g)$, where the marginal PDF $\widetilde{P}(\phi_g)$ corresponds to a $\beta$-function for $Z$ and the conditional PDF $P(\phi_s|\phi_g)$ corresponds to a $\delta$-function. Using this approximation, turbulence-soot interactions are partially accounted for through the sub-grid fluctuations of $Z$. As in the case of other species characterized by slow-time evolution, the soot cluster source term is splited into a production term   and a consumption term.
The latter is linearized as a function of $\widetilde{Y}_{s,j}^{clt}$ to prevent nonphysical soot consumption when $\widetilde{Y}_{s,j}^{clt}$ = 0. A detailed description of the model and approximations can be found in~\cite{kalbhor2023}.

\section{Results and discussion} \addvspace{10pt}

Validation and analysis of results are divided into three main subsections. First, gas-phase results are presented to discuss the predicting capabilities of the model to recover the flow field and flame shape.
Second, soot results are introduced to analyse the influence of air dilution on soot formation. Third, the predicted distribution of particle sizes along the burner are discussed. Numerical results are compared with experimental data for the gas-phase and soot quantities.

\subsection{Gas-phase analysis} \addvspace{10pt} 

\begin{figure}[ht!]
\centering
\includegraphics[width=0.9\linewidth]{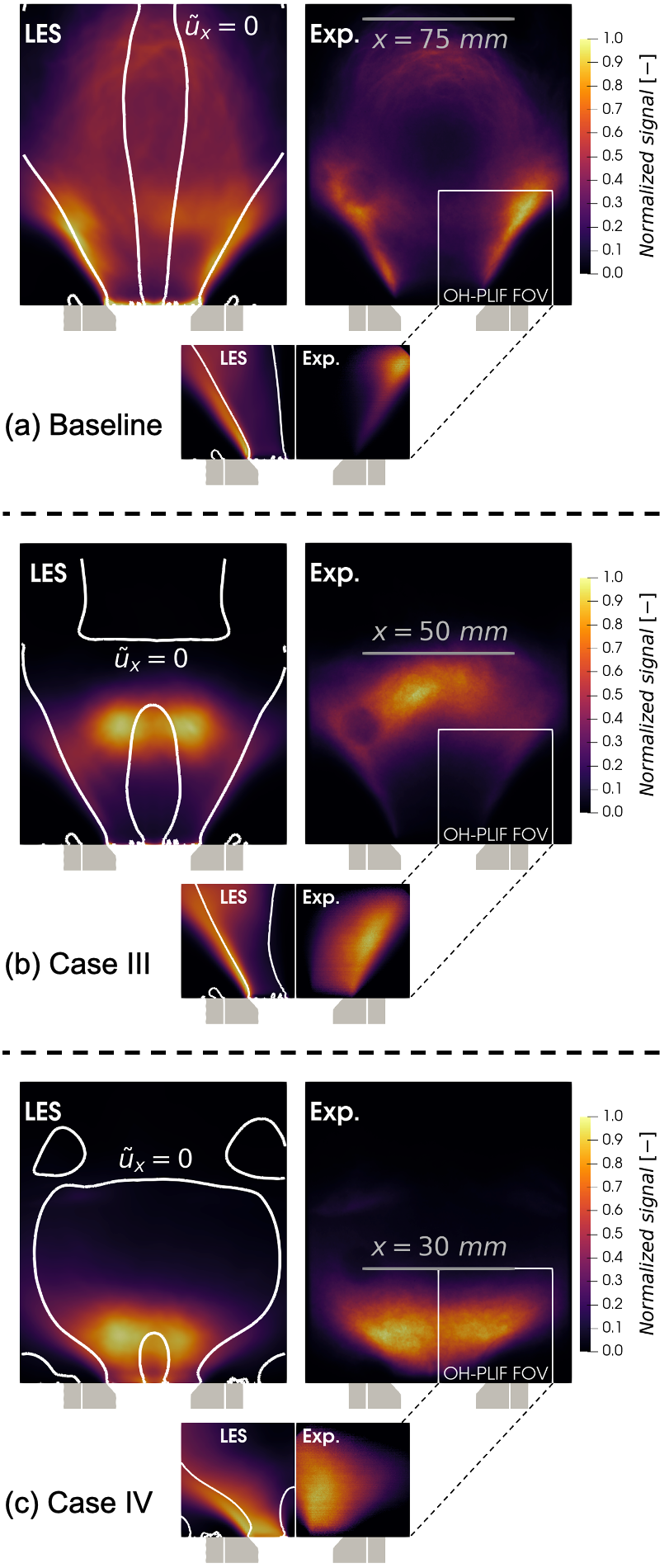}
\caption{Time-averaged normalized LOS signal: heat release rate for LES and \ce{OH^*} chemiluminescence for experiments \cite{elhelou2021} (FoV: 70$\times$\SI{79}{mm}). Time-averaged normalized planar signal: $\tilde{Y}_\mathrm{OH}$ for LES and OH-PLIF for experiments \cite{elhelou2021} (FoV: 30$\times$\SI{30}{mm}). White isocontour $\tilde{u}_{x} = 0$.}
\label{fig:HRR_YOH}
\end{figure}

Given the strong correlation between mixing, combustion and soot production processes, it is important to assess the predicting capabilities of the flamelet model to recover the flame shape in this configuration.
Normalized line-of-sight (LOS) signals are displayed in Fig.~\ref{fig:HRR_YOH} based on \ce{OH^*} chemiluminescence \cite{elhelou2021} for the experiments and heat release rate (HRR) of the LES. The field of view (FoV) in the figure is 70$\times$\SI{79}{mm}.

The mean flame topology for the three operating conditions is shown in Fig.~\ref{fig:HRR_YOH}.
In line with experimental observations, the flame for the Baseline case exhibits a canonical shape with no signal at the outer recirculation zone (ORZ).
By adding 20\% dilution air (Case III), see Fig.~\ref{fig:HRR_YOH}(b), the flame retains a similar shape, but peak values shift from the inner share layer (ISL) to the CRZ around the chamber's axis. With 40\% dilution air (Case IV), the flame topology differs significantly
(see Fig.~\ref{fig:HRR_YOH}(c)). Under this condition, the zone with peak values is shifted towards the bluff body spanning the full width of the CRZ.
The flame becomes unstable in this case, exhibiting strong oscillations and complex dynamics due to the large volumetric flow rates introduced through the dilution jets.
Regarding flame shape, one of the most distinctive features is the reduction in flame length. 
This shortening, resulting from the interaction with the air dilution jets injected at \SI{47}{mm}, is well captured by the LES as seen from the normalized LOS HRR signal.

The zoomed plots in Fig.~\ref{fig:HRR_YOH} compare predicted mass fraction of \ce{OH} from the LES with the measured \ce{OH} planar laser induced fluorescence (PLIF) from \cite{elhelou2021}. The Baseline condition shows a narrow distribution along the ISL. In Case III and Case IV, the \ce{OH} distribution widens due to dilution jets, affecting both spatial distribution and flame stabilization near the bluff body. The LES results are consistent with the experimental findings reported in~\cite{elhelou2021}.
\ce{OH}-PLIF measurements reveal
occasional flame front lift-off for the Baseline condition. This transient event is not captured by the flamelet approach, which may explain 
the discrepancy in the location of peak normalized \ce{OH} signal
along the ISL.
This edge flame propagation is a complex partially premixed combustion phenomenon which needs further model development to be captured using tabulated chemistry \cite{both2022}.
For Case IV, the flame front stabilization at the bluff body is well predicted as also found experimentally \cite{elhelou2021}. These transient flame dynamics have a low impact on soot formation in the primary zone as soot is mainly produced in high-temperature regions over long residence time, so the present formulation is retained for this study.

Overall, time-averaged LOS and planar results for the three conditions under study indicate 
that the modelling approach captures the key features of the different mean flame topologies. With no mixing/velocity experimental data available, the good correlation between LES and experimental results for both \ce{OH^*} and \ce{OH} suggests a reasonable prediction of the mixing process, hence mixture fraction field, enabling the study of soot formation and particle dynamics.

\subsection{Soot volume fraction distribution} \addvspace{10pt}

\begin{figure}[t!]
\centering
\includegraphics[width=0.95\linewidth]{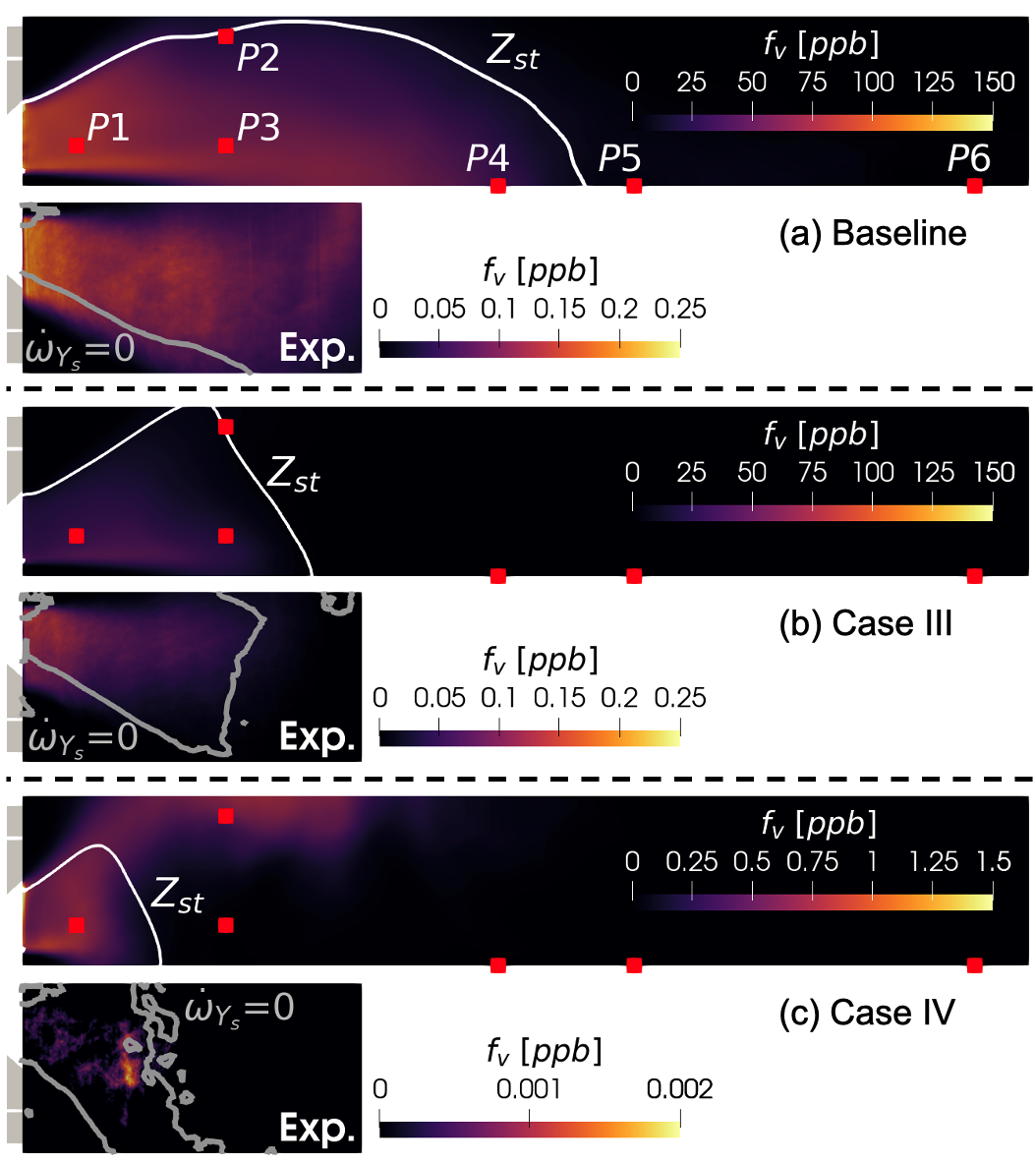}
\caption{Experiments: time-averaged $f_v$ field from LII \cite{defalco2021}. LES: time-averaged $f_v$ field (FoV: 50$\times$\SI{148}{mm}). Isocontours gray:
$\overline{\dot{\omega}}_{Y_s} = 0$ , white: $\tilde{Z}$ = $Z_{st}$.}
\label{fig:fv_field}
\end{figure}

\begin{figure}[t!]
\centering
\includegraphics[width=0.95\linewidth]{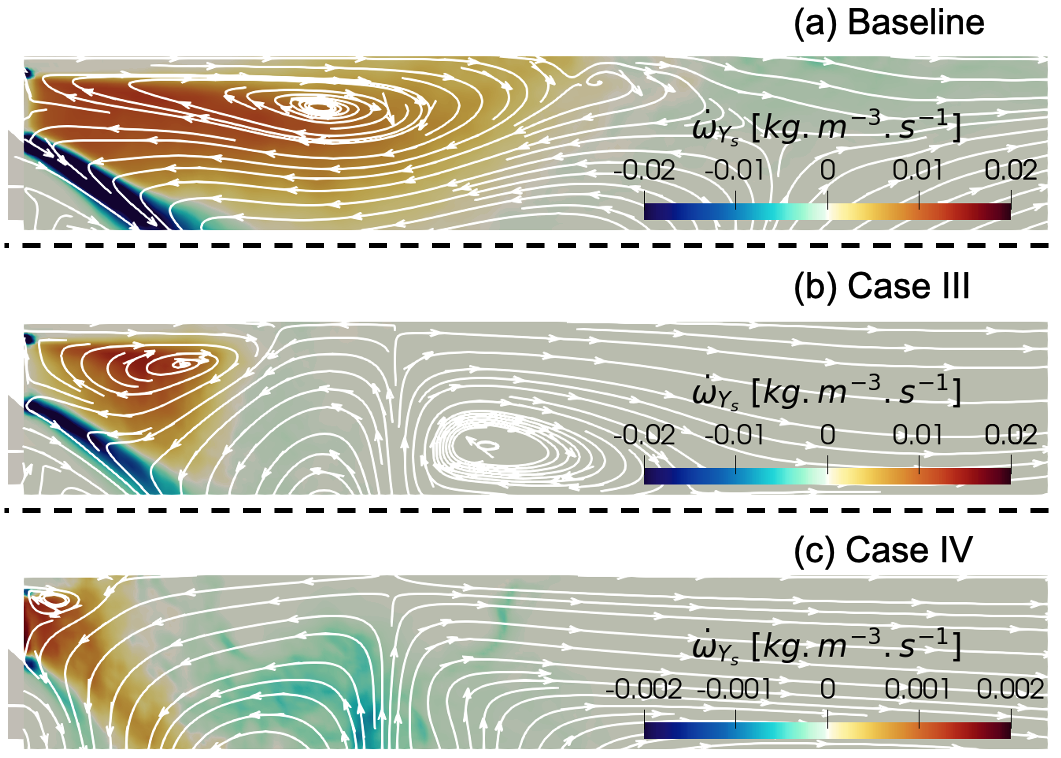}
\caption{LES time-averaged $\overline{\dot{\omega}}_{Y_s}$ with streamlines (FoV: 50$\times$\SI{148}{mm}).}
\label{fig:fv_src}
\end{figure}

In order to assess the feasibility of the modelling approach to predict the influence of air dilution on soot formation, the time-averaged $f_v$ field and the net soot mass fraction source term are shown for the three cases in Fig.~\ref{fig:fv_field} and Fig.~\ref{fig:fv_src}, respectively. The time-averaged laser induced incandescence (LII) measurements from \cite{defalco2021} are also included to compare the spatial distribution of $f_v$ between the LES and experiments. Without secondary air dilution (Fig.~\ref{fig:fv_field} (a)), the largest concentration of soot is found in the vicinity of the bluff body as seen from the LII and the LES. The predicted shape of the soot field is also in good agreement with the measurement as evidenced by the correlation between the LII signal and the isocontour of $\overline{\dot{\omega}}_{Y_s}$ = 0, which delimits the zones at which (on average) soot is produced ($\overline{\dot{\omega}}_{Y_s}\!>\!$ 0) and oxidized ($\overline{\dot{\omega}}_{Y_s}\!<\!$ 0) as seen in Fig.~\ref{fig:fv_src}. Predicting the spatial distribution of $f_v$ is challenging given the complexity of the soot production process. State-of-the-art results \cite{gkantonas2020} revealed an underestimation of soot around the bluff body (due to underestimated contribution of the smallest particles) in comparison with the most sooting region located around \SI{50}{mm} downstream. 
In this regard, the sectional soot model with the clustering approach in this work successfully 
recovers the spatial distribution described experimentally with a reduced computational cost.

\begin{figure}[h!]
\centering
\includegraphics[width=0.9\linewidth]{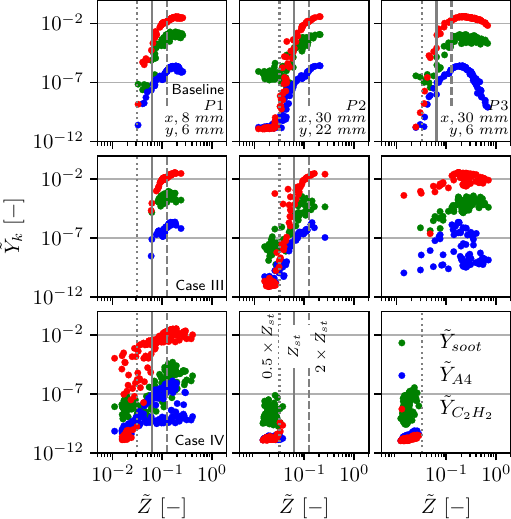}
\caption{LES instantaneous scatter of mixture fraction and mass fraction of soot (green), \ce{A}4 (blue), and \ce{C2H2} (red) at different locations for investigated cases.
Dotted line: $0.5Z_{st}$, solid line: $Z_{st}$, and dashed line: $2Z_{st}$.}
\label{fig:scatter_P1P3}
\end{figure}

Additional results are collected at probe locations (highlighted with red square markers in Fig.~\ref{fig:fv_field}). Probes P1, P2, and P3 characterize the most sooting zone of the flame upstream the location at which air dilution is introduced ($x$ = \SI{47}{mm}). Time-average $f_v$ results show a decrease from P1 (\SI{83.5}{ppb}) to P3 (\SI{62.8}{ppb}). Several factors contribute to the drop in $f_v$ along the axial direction. First, rich mixtures (more favorable for soot formation) are predominant closer to the bluff body. Instantaneous mass fraction and mixture fraction scatters in Fig.~\ref{fig:scatter_P1P3} show that significant amounts (larger than \num{1e-7}) of soot (in green), \ce{A}4 (in blue), and \ce{C2H2} (in red) are mostly found at $\tilde{Z}\!>\! Z_{st}$ at P1. Further downstream at P3, the mixture fraction range extends  to even richer conditions $\tilde{Z}\!>\!2Z_{st}$, limiting soot production as noticeable for \ce{A}4. Lastly, lean mixtures are more prominent with a considerable presence of points at $\tilde{Z}\!<\! 0.5Z_{st}$ at P2. Second, the streamlines drawn on the time-averaged $\overline{\dot{\omega}}_{Y_s}$ field illustrate flow recirculation in the soot-prone zone.
The longer residence times and recirculation favor soot production and accumulation in the region closer to the bluff body. 
The highest values of $\overline{\dot{\omega}}_{Y_s}$ align with the recirculation vortex,
with P1 and P3 showing that the mixing state at this location is favorable for the formation of \ce{A}4 and \ce{C2H2} (key species for nucleation and HACA surface growth, respectively), hence soot. In contrast, P2 near the isocontour of $Z_{st}$ lies within the oxidation zone favored by air entrainment.

With 20\% air dilution (Fig.~\ref{fig:fv_field} (b)), there is a general reduction of the amount of soot produced and the extension of the $f_v$ field as a result of the interaction with the two counter-rotating vortex created by the air dilution jets. The isocontour of $\overline{\dot{\omega}}_{Y_s} = 0$ is in good agreement with the experimental LII signal indicating that the modelling approach reproduces well the interaction with the air dilution jets. Similar observations from the Baseline case are also valid for Case III. More soot is predicted near the bluff body at P1 (\SI{32.4}{ppb}) compared to the downstream location P3 (\SI{17.9}{ppb}). The enhanced mixing promoted by the air dilution jets leads to  spreading of the reacting front for various quantities including $\tilde{Z}$, soot, \ce{A}4, and \ce{C2H2}. 
The streamlines on the $\overline{\dot{\omega}}_{Y_s}$ field in Fig.~\ref{fig:fv_src} also show that higher concentrations of soot closer to the bluff body (P1 and P3) are favored by the vortex that coincides with the highest values of $\overline{\dot{\omega}}_{Y_s}$, while lower soot concentrations at P2 correlate with leaner mixtures and strong soot oxidation near the $Z_{st}$ isocontour. With 40\% air dilution, the extension of the sooting region is further reduced (both experimentally and numerically) compared to the Baseline and Case III. The drop in $f_v$ is linked to enhanced mixing by air dilution jets. 
The increased amount of secondary air dilution leads to reduced residence times in the CRZ impacting soot growth, supported by soot oxidation which also plays a significant role in mitigating soot formation by oxidizing larger soot particles.
The scatter in Fig.~\ref{fig:scatter_P1P3} 
reveals more spreading (as compared to the Baseline and Case III) along the axis at the most sooting region (P1). At P2 and P3, most of the mixture drops below $0.5 Z_{st}$, limiting soot production.

Overall, the proposed LES FGM-CDSM approach effectively reproduces the trends and spatial distribution of $f_v$ for the three cases.
In terms of magnitude, the addition of 20\% dilution limits soot production keeping the same order of magnitude as the Baseline condition (0--\SI{150}{ppb}).
With 40\% air dilution, the drop in $f_v$ is more pronounced differing by at least one order of magnitude (0--\SI{1.5}{ppb}) from the Baseline and Case III. This trend aligns with the experimental measurements for which the range of $f_v$ remains between 0--\SI{0.25}{ppb} for the Baseline and Case III and drops to 0--\SI{0.002}{ppb} for Case IV. Validating the magnitude of $f_v$ from the LES is  challenging given the complexity of quantitative measurements in turbulent flames. The comprehensive set of measurements from \cite{defalco2021} indicates that the difference in $f_v$ across several techniques might lay within a factor ten in the case of LII and {\em in situ} probe measurements. This is partly because LII measurements are intrinsically biased towards larger particles and may not capture the full amount of soot produced \cite{defalco2021}. 

\begin{figure}[h!]
\centering
\includegraphics[width=0.8\linewidth]{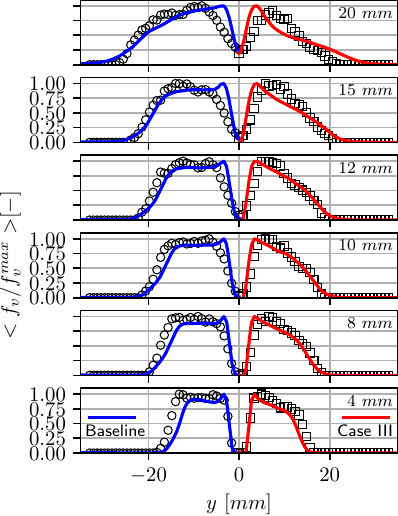}
\caption{Predicted (color lines) and measured (markers) time-averaged normalized $f_v$ radial profiles. Baseline case (blue, $\circ$), Case III (red, $\triangledown$). Experimental results \cite{defalco2021}.}
\label{fig:fv_profiles}
\end{figure}

Validation of the spatial distribution of soot is complemented with the comparison of their time-averaged normalized radial profiles in Fig.~\ref{fig:fv_profiles} for the Baseline (left) and Case III (right).
In general, the predicted shape of the profiles correlates well with the experimental data.
Without secondary air dilution, the top-hat-like profiles widen with increased axial distance until $x$ = \SI{20}{mm} where the profile is not flat beyond $y\!>\!$ \SI{15}{mm}. At this location, soot oxidation becomes more predominant.
For Case III, the transition from the top-hat-like shape is faster as soot oxidation is enhanced through the introduction of air dilution.

\subsection{Analysis of soot particle size distribution} \addvspace{10pt}

\begin{figure}[h!]
\centering
\includegraphics[width=0.9\linewidth]{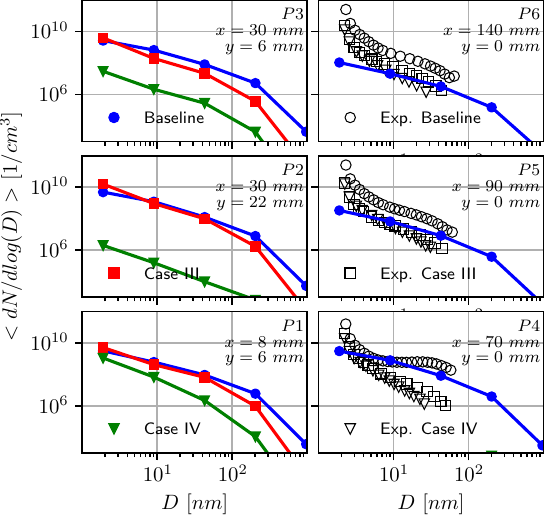}
\caption{Predicted (color lines) and measured (black markers) number PSD at different locations. Baseline case ($\circ$), Case III ($\triangledown$), and Case IV ($\Box$). Experimental data from \cite{defalco2021}.}
\label{fig:PSD}
\end{figure}

The use of the LES FGM-CDSM approach facilitates the information on soot 
PSD. The predicted PSD, retrieved from the six clustered sections, at all locations highlighted with red
markers in Fig.~\ref{fig:fv_field} (P1 to P6) is shown in Fig.~\ref{fig:PSD} along with the experimental SMPS measurements from \cite{defalco2021} at P4, P5, and P6. 
Qualitatively, PSD predictions between 70--\SI{140}{mm} correlate well with measurements showing a unimodal distribution of particles.
Experimentally, negligible changes between Case III and Case IV were reported at any of the probe locations, which may be attributed to the insensitivity of measurements downstream of the air dilution jets \cite{defalco2021}. Numerically, the number of particles is underpredicted and the range falls outside the range displayed in Fig.~\ref{fig:PSD}. PSD predictions at locations upstream from the air dilution jets (P1, P2, and P3) reveal that, for the Baseline case and Case III, the PSD do not change considerably across the combustor. This observation holds for Case IV at P1 which lays within the most sooting zone of the flame, see Fig.~\ref{fig:fv_field} (c). At P2 and P3, the number of particles is considerably reduced due to the 
increased soot oxidation in the primary zone controlled by the dilution jets.

\begin{figure}[h!]
\centering
\includegraphics[width=0.8\linewidth]{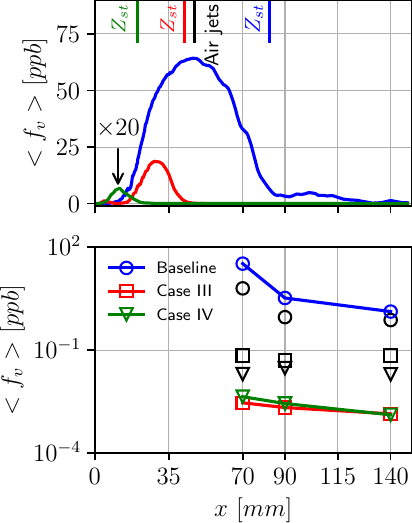}
\caption{Predicted (color lines) and measured (black markers) $f_v$ along the combustor's axis. Baseline case ($\circ$), Case III ($\triangledown$), and Case IV ($\Box$). Experimental data from \cite{defalco2021}.}
\label{fig:fv_axis}
\end{figure}

The predicted drop in PSD is elucidated by comparing the predicted and measured mean $f_v$ along the combustor's axis in Fig.~\ref{fig:fv_axis}, where vertical color lines denote the location of mean $Z_{st}$ for each of the cases. Between 70--\SI{140}{mm}, the trend of decreasing $f_v$ with increased axial distance observed experimentally aligns with the numerical results for the Baseline condition. Predictions reveal that the drop in $f_v$ from P4 to P5 correlates with the transition (on average) from rich to lean mixtures. For Case III and Case IV, the amount of soot is predicted to be several orders of magnitude lower with no considerable changes between cases as also found experimentally.
Numerical predictions indicate that under RQL operation, soot is limited through enhanced mixing (leaner mixtures observed in Fig.~\ref{fig:scatter_P1P3}) and a more constrained soot-prone zone. By looking at the predicted mean $f_v$ profiles upstream \SI{70}{mm} (dashed lines), it is seen how the primary zone is constrained for Case III with $Z_{st}$ found on the axis at \SI{42.6}{mm} as compared to \SI{82.7}{mm} for the Baseline. With increased air dilution, the primary zone in Case IV is further constrained  with $Z_{st}$ on the axis at \SI{20.3}{mm}. Leaner mixtures and the shift (on the axis) of $Z_{st}$ upstream the air dilution jets are responsible for the reduction in soot formation seen for Case III and Case IV.

The LES results demonstrate that the flamelet approach combined with DSM effectively captures the soot mitigation mechanism promoted by the dilution jets. Moreover, qualitatively good predictions of the PSD for the Baseline condition are reported. Nevertheless, the comparison with the experiments show an overprediction of the oxidation rates at $x\!>\!$ \SI{70}{mm}. In such conditions, the number of particles reduces drastically and it is more difficult to capture the correct PSD. This overestimation could be related to the univariate description of the soot particles. A more complex formulation of the soot model (accounting for the aggregate structure of soot) could overcome this limitation. It could potentially provide a better description of surface-related phenomena (including soot oxidation) and an improved prediction of the PSD, while allowing a more consistent comparison with the SMPS measurements. However, this is left for future work.

\section{Conclusions} \addvspace{10pt}

Soot formation in a lab-scale RQL swirl combustor was numerically studied using an LES FGM-CDSM approach.
This work extended the application of the LES FGM-DSM approach \cite{kalbhor2024} to the C-RQL combustor, while further validating the FGM-CDSM approach \cite{kalbhor2023} for turbulent conditions. LES predictions provide insight on soot reduction under RQL operation contributing to a better understanding of the trends found experimentally \cite{elhelou2021,defalco2021}.

Results of the mean flame shape along with predictions of reduced flame length and widening of the reaction zone (resulting from the interaction of the flame with air dilution jets) were shown to be in good agreement with the experimental data.
The good correlation of the predicted $f_v$ field with LII measurements is a significant improvement over the state-of-the-art results in the literature. With air dilution, soot production is limited by favored soot oxidation linked to the contraction of the flame size (on the axis $Z_{st}$ moves upstream the air dilution jets) and stronger mixing, hence leaner mixtures.
Qualitatively, the comparison of SMPS PSD with LES results are in acceptable agreement with the unimodal distribution predicted by the LES. Accounting for the aggregate structure of soot could potentially improve the prediction of the PSD for cases with air dilution and is currently being explored.

The analysis of results evidences that the LES FGM-CDSM approach offers a good trade-off between accuracy and computational cost to study soot production in a combustor model representative of a gas turbine. 

\acknowledgement{Acknowledgments} \addvspace{10pt}

This project received funding from the Center of Excellence in Combustion project (grant agreement No 952181), the AHEAD PID2020-118387RB-C33 and SAFLOW TED2021-131618B-C2 projects (Ministerio de Ciencia e Innovaci\'on). LP acknowledges the Margarita Salas grant (Ministerio de Universidades, EU-Next Generation EU) and DM acknowledges the grant Ramón y Cajal RYC2021-034654-I.
The authors acknowledge computer resources from the RES (IM-2023-2-0011 and IM-2023-3-0013).

% -------------------------------------------------------------------- %
% -------------------------------------------------------------------- %
% -------------------------------------------------------------------- %

 \footnotesize
 \baselineskip 9pt

% -------------------------------------------------------------------- %
% -------------------------------------------------------------------- %
% -------------------------------------------------------------------- %

\bibliographystyle{pci}
\bibliography{refs}

% -------------------------------------------------------------------- %
% -------------------------------------------------------------------- %
% -------------------------------------------------------------------- %

\newpage

\small
\baselineskip 10pt

% -------------------------------------------------------------------- %
% -------------------------------------------------------------------- %
% -------------------------------------------------------------------- %

% -------------------------------------------------------------------- %
% -------------------------------------------------------------------- %
% -------------------------------------------------------------------- %

\twocolumn[\begin{@twocolumnfalse}

{\bf Novelty and Significance Statement}
\vspace{10pt}

The novelty of this research lays on: (1) investigating soot production in the Cambridge rich-quench-lean combustor through large-eddy simulations comprising tabulated chemistry (FGM) with tabulated soot source terms using a clustered discrete sectional method (CDSM) and (2) providing insight to better understand the experimental trends. Numerically, it was found that the reduction of soot formation in the cases with air dilution is driven by favored oxidation (facilitated by the contraction of the flame size) and by stronger mixing. This work is significant because the effect of air dilution on soot production is characterized and validated with experimental data and because it demonstrates that the proposed LES FGM-CDSM methodology allows for a detailed investigation of soot formation and particle dynamics in a practical combustor at reduced computational cost.
Also, the soot volume fraction spatial distribution is well reproduced which constitutes an improvement respect to state-of-the-art results from the literature. 

\vspace{20pt} 

{\bf Author Contributions}
\vspace{10pt}

\begin{itemize}
  \item{L. P.: Conducted research, analyzed data, wrote the paper}

  \item{A. K.: Developed and implemented code, wrote the paper}

  \item{D. M.: Designed research, developed and implemented code, wrote the paper}

  \item{J. v. O.: Designed research, revised the paper}
\end{itemize}

\end{@twocolumnfalse}] 

% -------------------------------------------------------------------- %
% -------------------------------------------------------------------- %
% -------------------------------------------------------------------- %

\end{document}